\documentclass[sigconf,dvipsnames]{acmart}
\AtBeginDocument{%
  \providecommand\BibTeX{{%
    \normalfont B\kern-0.5em{\scshape i\kern-0.25em b}\kern-0.8em\TeX}}}

\usepackage{xcolor}
\usepackage{framed}

\usepackage{dblfloatfix}

\usepackage{array}
\usepackage[most]{tcolorbox}

\renewcommand\footnotetextcopyrightpermission[1]{} % removes footnote with conference information in first column

\makeatletter
\newcommand{\interviewquote}[2]{
 \def\FrameCommand{%
    \hspace{0pt}%
    {\color{MidnightBlue}\vrule width 1.5pt}%
    {\color{white}\vrule width 4pt}%
    \colorbox{white}
  }%
  \MakeFramed{\advance\hsize-\width\FrameRestore}%
  \noindent\hspace{-4.55pt}% disable indenting first paragraph
  \footnotesize{"\emph{#1}" -{#2}}\vspace{0.5pt}\endMakeFramed%
}
\newcolumntype{P}[1]{>{\centering\arraybackslash}p{#1}}
\newcolumntype{M}[1]{>{\centering\arraybackslash}m{#1}}
\makeatother

\setcopyright{none}

\begin{document}

\title{Facing the Giant: a Grounded Theory Study of Decision-Making in Microservices Migrations}

\author{Hamdy Michael Ayas}
\email{ayas@chalmers.se}
\affiliation{%
  \institution{\textit{CSE Department} \\ 
  \textit{Chalmers $|$ University of Gothenburg}}
  \city{Gothenburg}
  \country{Sweden}
}
\author{Philipp Leitner}
\email{philipp.leitner@chalmers.se}
\affiliation{%
  \institution{\textit{CSE Department} \\ 
  \textit{Chalmers $|$ University of Gothenburg}}
  \city{Gothenburg}
  \country{Sweden}
}
\author{Regina Hebig}
\email{hebig@chalmers.se}
\affiliation{%
  \institution{\textit{CSE Department} \\ 
  \textit{Chalmers $|$ University of Gothenburg}}
  \city{Gothenburg}
  \country{Sweden}
}

\renewcommand{\shortauthors}{Michael Ayas, et al.}

\begin{abstract}
\textbf{Background:} Microservices migrations are challenging and expensive projects with many decisions that need to be made in a multitude of dimensions. 
Existing research tends to focus on technical issues and decisions (e.g., how to split services). Equally important organizational or business issues and their relations with technical aspects often remain out of scope or on a high level of abstraction. \\
\textbf{Aims:} In this study, we aim to holistically chart the decision-making that happens on all dimensions of a migration project towards microservices (including, but not limited to, the technical dimension). \\ 
\textbf{Method:} We investigate 16 different migration cases in a grounded theory interview study, with 19 participants that recently migrated towards microservices. 
This study strongly focuses on the human aspects of a migration, through stakeholders and their decisions. \\
\textbf{Results:} We identify 3 decision-making processes consisting of 22 decision-points and their alternative options. The decision-points are related to creating stakeholder engagement and assessing feasibility, technical implementation, and organizational restructuring. \\
\textbf{Conclusions:} Our study provides an initial theory of decision-making in migrations to microservices. It also outfits practitioners with a roadmap of which decisions they should be prepared to make and at which point in the migration.
\end{abstract}




\maketitle
\pagestyle{plain} % removes running headers

\section{Introduction}

% Background/intro of study
Organizations in many industries are increasingly adopting microservices technologies to design, develop, test and maintain software systems.
Development with microservices can happen in new software systems, but more commonly an existing system needs to be migrated to a microservices-based architecture (MSA) \cite{Balalaie2018}. 
Therefore, migrations towards microservices are increasingly gaining popularity in both industry and academia~\cite{Hassan2020, SOLDANI2018215}.
Also, early research and documented work on microservices demonstrates the complexity in the nature of designing and developing microservices 
\cite{Zimmermann2017, newman2015building}.
Consequently, changing the software architecture and adopting a MSA is also a complex endeavour \cite{Fritzsch2019}. 
Crucially, architectural migrations are heavy in decision-making \cite{Carvalho2019}, either in an individual-level, team-level or organizational-level (i.e. company wide). Defining the decision-making processes of migration initiatives in software development organizations helps to better understand such transitions and improve their realization \cite{Badampudi2018, AUER2021106600}.

Substantial previous research has investigated questions surrounding microservices migrations~\cite{Balalaie2018, Fritzsch2019}. These works tend to focus on technical issues and decisions (e.g., how to split services~\cite{Gysel2016, Zhongshan2018}) and on the software architecture \cite{Taibi2017}. Also, there are studies on how to technically enact the migration (e.g., through program decomposition) \cite{fritzsch2018monolith}. However, such solutions are often not sufficient in supporting the decision-making of engineers in migrations \cite{Carvalho2019}. 
Furthermore, microservices migrations are transformative on organizations as a whole, and we have a lesser understanding on the non-technical aspects of migrations than the technical aspects. 
Research and best practices stemming from industry provide some comprehensive approaches on migrations, covering many aspects \cite{newman2019monolith, Balalaie2016}. However, there is a lack of approaches providing details on the operational choices that software development teams and organizations make in migrations. Also, there is room for the empirical understanding of migrations from engineers' point of view. 

Hence, the objective of this study is to holistically chart the decision-making processes that happen on all levels of a microservices migration project, inductively from empirical evidence. 
A strong emphasis is given on the multidimensional nature of migrations towards microservices, considering the business and organizational side, as well as the technical side. 
Our study approaches the topic with a strong focus on the human aspect of a migration, through stakeholders, their concerns and the decisions they make when migrating. 
We analyze 16 different cases of migrations towards microservices from 16 different organizations, via conducting a grounded theory based interview study with 19 developers. All developers have recently been part of a migration towards microservices.
The research questions we study are: 

\vspace{0.2cm}
\noindent \emph{\textbf{RQ1}: What is the decision-making process of organizations during a migration towards microservices?} 
\begin{description}
    \item \emph{\textbf{RQ1.1}: What are the decisions that organizations make?}
    \item \emph{\textbf{RQ1.2}: At what point in the migration are decisions made?}
\end{description}
\emph{\textbf{RQ2}: What alternative options do software development organizations choose in each of these decisions?} 
\vspace{0.2cm}

We construct an initial theory of decision-making in microservices migrations.
The decision-making in our theory takes place in an organizational level (i.e. collective decisions taken by the software development organization). The decision-making processes we construct consist of decision-points, typical options, and dependencies between them (e.g., follow-up decisions based on the outcome of a previous decision). We identify 22 decision-points with their potential options. These relate to the assessment of technical feasibility, the validation of the migration through a business case, the technical implementation, as well as the restructuring of the organization and its operations. Moreover, we distinguish two types of decision-points: (1) procedural decision-points (decisions about how to continue the migration project, i.e., which strategies to use), and (2) outcome decision-points (decisions about the MSA that shall result from the migration).

Through the identified decision-making processes, we demonstrate the choices that organizations make during the course of a migration, from its early stages. The contents of our identified decision-making processes complement existing knowledge on utilizing the benefits of microservices. We demonstrate that microservices are not only motivated by economic gains of efficiency, but also by gains in effectively and continuously delivering business value. 
Subsequently, we demonstrate a comprehensive view of migrations with decision-making not only on the technical dimension (e.g. developing microservices), but also the organizational. Meaning that we present how the software development organization is structured, for example with new ways of managing teams. 
Our focus is mainly on understanding the decisions that stakeholders (e.g. developers, managers) make during migrations and the situations they come across. We do not argue that we provide a single source of truth on how to migrate. In fact, we believe that such a thing does not exist due to the complexity and socio-technical nature of software development organizations migrating.

\section{Related Work}
With migrating to microservices, organizations aim to deconstruct their systems into a set of smaller, independent services \cite{Dragoni2017}. In principle, these services are decoupled from each other, with minimal dependencies between them \cite{newman2015building}. In addition, microservices are supposed to be individually owned by teams that are responsible to design them around business domains \cite{Zimmermann2017}. To achieve domain-driven design, teams are supported by design or architectural patterns that also help to achieve the benefits of microservices \cite{thones2015microservices}. Such patterns include organizing software systems around business capabilities, enabling automated deployment, facilitating intelligence in endpoints and decentralizing the control of programming languages and data \cite{DiFrancesco2019}.
Therefore, usually there is a large leap between a monolith and a microservices architecture and a migration / transition project is the crucial project taking the organization through the leap \cite{Hassan2020}.
Our research investigates this transition leap and we give a detailed focus on the underlying decision-making.

Organizations are increasingly migrating their software architectures by adopting microservices and researchers are increasingly investigating solutions for such transitions \cite{Hassan2020}. A substantial amount of previous research has investigated the area surrounding microservices migrations and their characteristics \cite{Balalaie2018, Taibi2017, Fritzsch2019}. In addition, there are solutions researched on how to technically enact the migration. These solutions include splitting a system, transforming the code of an application or identifying services \cite{fritzsch2018monolith}. These solutions often provide tools on how to identify and decompose or extract services, assuming a technical viewpoint on the migration \cite{Gysel2016}. This is not always ideal, as other aspects are neglected that usually come along with a migration, related to managing the entire change of an organization \cite{newman2019monolith}. Industry-based research uses influential work from the discipline of change management to demonstrate how to lead the change of a migration. For example, with the 8-steps model to transforming an organization \cite{kotter1995leading}. Hence, it could prove valuable to go even further in detail in such influential work, for example on specific activities on how to manage resistance to change through education, participation, facilitation, negotiation and coercion \cite{kotter1979choosing}. This will help to understand the ways for consistently achieving the strategic alignment needed to leverage the technology \cite{Venkatraman1999} of microservices in this case. 

Microservices enable many benefits in the development and operations of software systems \cite{Zdun2020}. For example, microservices enable the unprecedented possibility of dynamically scaling-up and scaling-down parts of an application \cite{dragoni2018}. Furthermore, the modular organization of systems with minimal dependencies, may offer improved maintainability \cite{Zdun2020} and organizational agility \cite{Zimmermann2017}. 
Also, characteristics of microservices such as infrastructure automation enable continuous delivery, improving operational efficiency \cite{singleton2016}. For example, the average size of development teams and the domain-specific redundancy between microservices can be decreased in comparison to monolith systems \cite{Mazlami2017}. 
Hence, the proposition of migrating to microservices enables substantial economic potential in efficiently developing and managing complex software systems \cite{singleton2016}.
However, the aspect of organizational and business effectiveness is equally important \cite{Venkatraman1986}, even though it is not studied as extensively in the context of microservices. 
Consequently, we combine with empirical evidence the aforementioned aspects, along with how organizations can justify their decisions around migrating.

Migrations are intensive in operational decisions \cite{Carvalho2019}. Just like every transformation initiative, migrating to microservices entails many risks to consider \cite{Lin2016}. For example, starting from a greenfield to migrate can be very expensive but re-factoring an established system is a long-lasting endeavour, influenced by a broad range of aspects \cite{fritzsch2018monolith}. In such cases, understanding the decision-making process is beneficial \cite{Badampudi2018}. Decision-making can be incorporated in all architectural views to facilitate our understanding of systems which is particularly valuable during change \cite{Kruchten2009}. Hence, it is not only needed to understand the criteria for decision-making in migrations towards microservices \cite{Carvalho2019}, but also to understand the decision-making processes (with their other constituent elements).

A key challenge in decision-making within software engineering is to frame decisions into well-structured, well-defined problems \cite{zannier2007model}. Then, according to \citeauthor{zannier2007model}, it is more likely to have a better understanding of the problem's constituent elements and to organize a rational process for addressing it. 
However, rational decision-making is often not realistic and assuming a deterministic decision-making process is problematic \cite{tang2017human}. 
Decision making of software engineers is dominated from rationality in well-defined, structured problems with known quality attributes and non-rational approaches (``naturalistic decision making'') are adopted in unstructured problems that fall out of existing experiences \cite{zannier2007model}. 
These limitations along with their implications are extensively studied in other fields, but only moderately studied in software engineering \cite{tang2017human}. Furthermore, with the dominating focus on defining software engineering processes, it is attempted to force the adoption of certain practices that are derived from the technical capabilities that are in place \cite{parnas2009document} and not from the humane particularities that an organization’s system of developing software has \cite{tang2017human}. 

Therefore, without having these humane particularities in the core of microservices migrations, it is challenging to have a realistic picture of microservices migrations. 
Our study addresses this by giving a strong emphasis on the socio-technical nature of microservices migrations, using dimensions described by \citeauthor{Baxter2011} \cite{Baxter2011}. Our comprehensive perspective takes place in multiple dimensions (technical and otherwise). This comes in contrast with most studies in state-of-the-art, that according to \citeauthor{Hassan2020} investigate migrations as a technical endeavor that needs a technical solution \cite{Hassan2020}. We do not only approach migrations as a technical endeavor, but also as an endeavor with a strong social and business aspect to it. Additionally, there is no other study to the best of our knowledge that studies microservices migrations in the light of decision-making regarding detailed activities of migrations, which is important if we consider how many decisions engineers need to make when migrating.

\section{Methodology}

Our primary research method is a grounded theory based interview study with practitioners who have recently participated in a microservice migration project. 
The adopted inductive approach allows us to understand the decision-making process that software development organizations and developers go through during a migration. We derive our theory by isolating the situational factors of decision-making that migrations entail.
The interview guide used in the semi-structured interviews can be found in our replication package~\cite{michael_ayas_hamdy_2021_5060129}. We omit interview transcripts from the replication package to preserve interviewee privacy and protect potential commercial interests of our interviewee's employers.

\subsection{Interviews analysis}
Our interview analysis is based on techniques from Grounded Theory (GT)~\cite{charmaz2014constructing}. Specifically, we used coding, memoing, sorting, constant comparison and theoretical saturation to develop our theory. According to literature for GT in software engineering research, 
we cannot claim to use the classic GT method, but instead constructivist GT \cite{Stol2016}. A reason for this choice is that we had significant previous exposure to literature prior to the study. The previous exposure contributed to the alignment of some of our themes with both, previous research~\cite{Balalaie2018, Hassan2020} as well as common practitioner guidance on best practices for microservices migration \cite{newman2019monolith}. 

\paragraph*{Study Design:}
In accordance to constructivist GT, we started with an initial research question that evolved throughout the study \cite{Stol2016}. The initial research question was inspired from conversations with practitioners undergoing microservices migrations and supported by literature. Then, we further specified and broke down the initial research question into multiple ones that could be addressed based on the data analysis.
To conduct the interviews we used a semi-structured interview guide, which we constructed based on our initial research questions. However, we gave participants significant freedom in describing their migration experiences.
We chose this semi-structured guide, in order to achieve an approach with critical realism and identify the objective realities of microservices migrations, as they were experienced from the interviewees. 

\paragraph*{Participants:}
We recruited software developers that are diverse in terms of experience, age, roles and domain. This allowed us to obtain an objective view on microservices migrations and a sufficient number of participants to perform GT. To do so, we relied on purposive sampling~\cite{baltes:20} and our personal network (e.g., through current and previous projects, colleagues, or students). Furthermore, through snowballing, we asked interviewees to connect us further to other potential participants. We used a saturation approach~\cite{charmaz2014constructing} where we continued inviting participants in parallel to data analysis as long as new insights were gained. Our acceptance criteria for selecting interviewees to contact were \emph{(a)} software engineering professionals (not students) who \emph{(b)} have participated in or have closely observed a microservices migration project in their professional work. An overview of the participants is found in Table~\ref{tab:interviewees}. We have interviewed 19 professionals from 6 different countries (Cyprus, UAE, Germany, Romania, Sweden, The Netherlands), of which 18 were male and one female. Interviewers had on average 7.5 years of experience (ranging from 2 to 21) and they have worked at medium to large companies in twelve business domains. 

In addition, through studying the 16 cases from 16 organizations (also presented in Table \ref{tab:interviewees}), we are able to investigate different types of systems and organizational cultures. 
The migration cases are about systems delivered to external customers (e.g. Enterprise SaaS), in-house enterprise solutions for internal users and also Software Applications sold as a service (e.g. mobile app). All organizations have relatively intensive software development capabilities, but not all of them are software companies (i.e. companies whose main purpose is related to software development). Specifically, 10 organizations can be considered software companies (operating in different industries) and the rest (i.e. Org3, Org6, Org13, Org14, Org15, Org16) are organizations from more traditional industries.

% \vspace{-1.5em}
\begin{table*}[h!]
\scriptsize
\begin{tabular}{ c c c c c c c c c} 
% \hline
\rotatebox{45}{\emph{Case}} & \rotatebox{45}{\emph{Organization}} & \rotatebox{45}{\emph{Org. Size}} & \rotatebox{45}{\emph{Type of System}} & \rotatebox{45}{\emph{Industry}} & \rotatebox{45}{\emph{Interview}} & \rotatebox{45}{\emph{Role}} & \rotatebox{45}{\emph{Experience}} & \rotatebox{45}{\emph{Responsibility}} \\[0.1em]
\hline
% \\[0.0001em]
C1 & Org1 & 50 & Enterprise SaaS & Enterprise Software & I1 & Full stack developer & 2 (1) & Implementation \\[0.1em] 
C2 & Org2 & 4,000 & In-house software solution & Gaming & I2 & Software Engineer & 2 (2) & Design \& Implementation \\[0.1em] 
C1 & Org1 & 50 & Enterprise SaaS & Enterprise Software & I3 & Senior Team Leader & 12 (2) & Leading Migration Project \\[0.1em] 
C3 & Org3 & 36,000 & In-house software solution & Banking Systems & I4 & Software Engineer & 2 (1) & Design \& Implementation \\[0.1em] 
C4 & Org4 & 9,000 & Enterprise SaaS & Banking Software & I5 & Software Engineer & 19 (2) & Architectural Design \\[0.1em] 
C1 & Org1 & 50 & Enterprise SaaS & Enterprise Software & I6 & Software Engineer & 2 (1) & Implementation \\[0.1em]
C5 & Org5 & 3,000 & In-house software solution & Aviation Software & I7 & Software Engineer & 7 (2) & Design \& Implementation \\[0.1em] 
C6 & Org6 & 30,000 & Enterprise SaaS & Telecommunications & I8 & Software Developer & 3 (3) & Implementation \\[0.1em]
C7 & Org7 & 27,000 & In-house software solution & Enterprise Software & I9 & Computer Scientist & 5 (5) & Implementation \\[0.1em]
C8 & Org8 & 200,000 & Enterprise SaaS & Cloud Computing & I10 & Principal Software Engineer & 7 (4) & Leading Development \\[0.1em]
C9 & Org9 & 33,000 & Enterprise SaaS & Marketing Analytics & I11 & Software Engineer & 6 (3) & Design \& Implementation \\[0.1em] 
C10 & Org10 & 150 & Enterprise SaaS & Healthcare Software & I12 & Data Engineer & 6 (2) & Leading Development \\[0.1em] 
C11 & Org11 & 83,000 & Consulting for Enterprise SaaS & Cloud Computing & I13 & Senior Cloud Architect & 10 (5) & Leading Development \\[0.1em] 
C12 & Org12 & 50 & Software Application & Energy Software & I14 & Software Engineer & 4 (1.5) & Design \& Implementation \\[0.1em]
C12 & Org12 & 50 & Software Application & Energy Software & I15 & Software Architect & 4 (4) & Leading Development \\[0.1em]
C13 & Org13 & 30 & Enterprise SaaS & Logistics / Planning & I17 & Co-founder & 8 (5) & Leading Migration Project \\[0.1em] 
C14 & Org14 & 62,000 & Enterprise SaaS & Logistics / Planning & I16 & Software Architecture Consultant & 13 (4) & Leading Development \\[0.1em]
C15 & Org15 & 25 & Enterprise SaaS & Manufacturing & I18 & CTO & 10 (6) & Outsourcing Development \\[0.1em] 
C16 & Org16 & 1m & Enterprise SaaS & Manufacturing & I19 & Enterprise Architect & 21 (5) & Leading Development \\[0.1em]
% \hline
\end{tabular}
\caption{Interview participants and case organizations. Organizations size is reported in approximate numbers of full time employees. Experience is in years and values in brackets are on experience with microservices.}
\label{tab:interviewees}
\end{table*}
% \vspace{-2em}

\vspace{-0.1em}
\paragraph*{Protocol:}
We conducted interviews over a period of six months during 2020, and each interview took between 30 and 60 minutes. Interviews were conducted via video conferencing, allowing the participation of interviewees in such a geographical diversity. Prior to each interview, participants were asked to sign a consent form, allowing us to record the interview. Furthermore, participants were informed that they can drop out of the study at any point, which no interviewee made use of. We did not offer any financial rewards to participants.
% \vspace{0.1em}

\vspace{-0.1em}
\paragraph*{Analysis:}
In an ongoing process parallel to data collection, we performed initial, focused and theoretical coding based on the constructivist variant of GT~\cite{charmaz2014constructing, Stol2016}. In initial coding we fractured the data to find relevant statements. In focused coding, we aggregated and connected those excerpts into categories and themes until achieving saturation. In theoretical coding we specified the relationships of the connected categories and integrated them into a cohesive theory. Initial coding was conducted by the first author. All authors collaborated in focused coding in three card sorting and memoing sessions lasting three to four hours each. Theoretical coding took place by all authors in increments and weekly sessions of at least 1 hour each. 

Additionally, theoretical coding was partly supported by existing literature, in order to enhance the validity of our findings. This helped us to identify the objective realities of microservices migrations, without sacrificing epistemic relativism in our acquired understanding. Existing literature helped us to understand more comprehensively and in detail some statements of software developers during the interviews. The resulting findings are supported by statements from multiple participants. 

Finally, it is worth mentioning that there were three predominant ways used in identifying decisions in our interview material. The first was identifying interviewees' mentions of occasions that they had to make a choice from different alternatives. The second way was when interviewees seemed unsure about a choice they had made and discussed the rationale behind it, or choose two options simultaneously. The third way was when we identified different courses of actions taken from different interviewees for the same task at hand. A common denominator of all different ways was the attempt of interviewees to define to the best of their ability different challenges or problems they had to solve. This comes in line with how decisions are described in software engineering literature \cite{zannier2007model}.
% \vspace{-0.5em}

\subsection{Threats to Validity}
Some threats that are inherent in our chosen study methodology remain, which readers should take into consideration.

\vspace{-0.1em}
\paragraph*{External Validity}
First, we cannot claim representativeness of our study demographics for the software industry in general, as the study population has been sampled through our personal network and using a voluntary, opt-in procedure. To mitigate this threat, we selected migration cases covering software development organizations of different sizes, in different domains, and in different geographical regions. In addition, we selected interviewees with different levels of experience and also covering many roles and responsibilities. 

\vspace{-0.5em}
\paragraph*{Internal Validity}
In terms of internal validity, a threat is that we were pre-exposed to existing research as well as practitioner-focused guidance on how to conduct microservice migrations due to our previous interest in the field. This may have biased our interview design, and may have led that some decision-points (e.g., those not discussed in earlier work) may have been given less prominence or judged as unimportant during analysis. A limitation of our qualitative study design is that we cannot claim that the identified list of decision-points and alternatives is necessarily complete. Hence, the results we present here shall be seen as an initial theory of MSA migration only, with the potential for later extension.

In addition, it is possible that interviewees communicate slightly different behaviors from the reality. This could lead to some incorrect perceptions of the migration. In addition, it combines with the threat that some interviewees started describing the practices or actions they were supposed to do, rather than those actually taking place. To mitigate these threats, first we initiated the interviews with informally meeting each other before asking questions. Also, we designed interview questions in a way that allowed us to discuss what actually took place in migrations. Specifically, we did not only ask about practices followed, but also about deviations from practices and challenges or problems that appeared along the way \cite{michael_ayas_hamdy_2021_5060129}.
However, we were as non-intrusive as possible when interviewees were describing their migrations. 
In addition, we further confirmed in some cases some details from the transcripts. 

\section{Results}

Our study shows that migrations towards microservices entail a multitude of decisions on different dimensions and of different types. We broadly distinguish between the \emph{decision-making process of creating engagement}, which then influences the \emph{decision-making process on the technical dimension} and the \emph{decision-making process on the organizational dimension}. On all dimensions, practitioners encounter two types of decisions that take place in certain points of time: \emph{procedural decision-points} (i.e., points during the migration where they need to decide how to continue the migration process) and \emph{outcome decisions-points} (e.g.,  points during the migration where the team needs to agree on specific architectures or technologies as outcome of the migration process). Both types of decision-points are only infrequently restricting practitioners to make a commitment to a single approach or technology --- instead, we observe that practitioners often choose to follow multiple, complementary, approaches.

Additionally, even though we observed an iterative nature in migrations, we present the decisions in the temporal order that are taken during a potential isolated increment. This way the pragmatic reality is showcased, in which software development organizations cannot revert a decision taken in the past, even though they can revisit its contents. 
Moreover, the order of decisions is derived from aggregating the experiences of interviewees, from what they described. 
We now discuss decisions on creating engagement as well as organizational and technical decisions. We also discuss their dependencies and implications for migration projects. We generally focus on the "flow" of decision-points and common options. Note that providing an exhaustive list of all possible options for every decision is neither feasible nor the scope of this article.

% \vspace{-1cm}

\subsection{Decisions on Creating Engagement}

Decisions on creating engagement primarily relate to whether a migration is an endeavour that the organization wants to pursue in the first place. This requires evaluating technical feasibility as well as creating engagement with all key stakeholders. We identify key stakeholders as top management, middle management and operational personnel. Top management's buy-in is essential, as they act as the funding agency of the migration. Middle management is commonly the source of knowledge about how cost reductions or profit increases can be enacted in practice. Finally, operational personnel (including software developers and architects) engineer the new architecture, and are responsible for development and delivery. 

We observe that the studied cases of migration projects include decisions that address the concerns of all three of these stakeholder groups. Firstly, technical feasibility can be evaluated (predominantly to convince the operational and middle management stakeholders). Then follows the construction of convincing business cases for top management.
Therefore, creating management buy-in predominantly entails establishing a business case describing how microservices lead to reduced costs and/or increased profits (e.g. from a better product or service) for a strategic (and often very expensive) migration project. It should be noted that this process of creating and evaluating buy-in is not a one-time procedure -- instead, many investigated migration projects continuously (re-)evaluate if microservices are still the right fit for their project or company.
\interviewquote{the microservice approach allows an iterative growth [...] we start [...] a very shallow data inventory, then I tried to add applications on it, then I tried to drive the customer to say 'Okay' for the next application}{I18}

\subsubsection{Technical Feasibility and Exploration of Opportunities.}
\begin{figure}[h!]
  \centering
  \includegraphics[width=\linewidth]{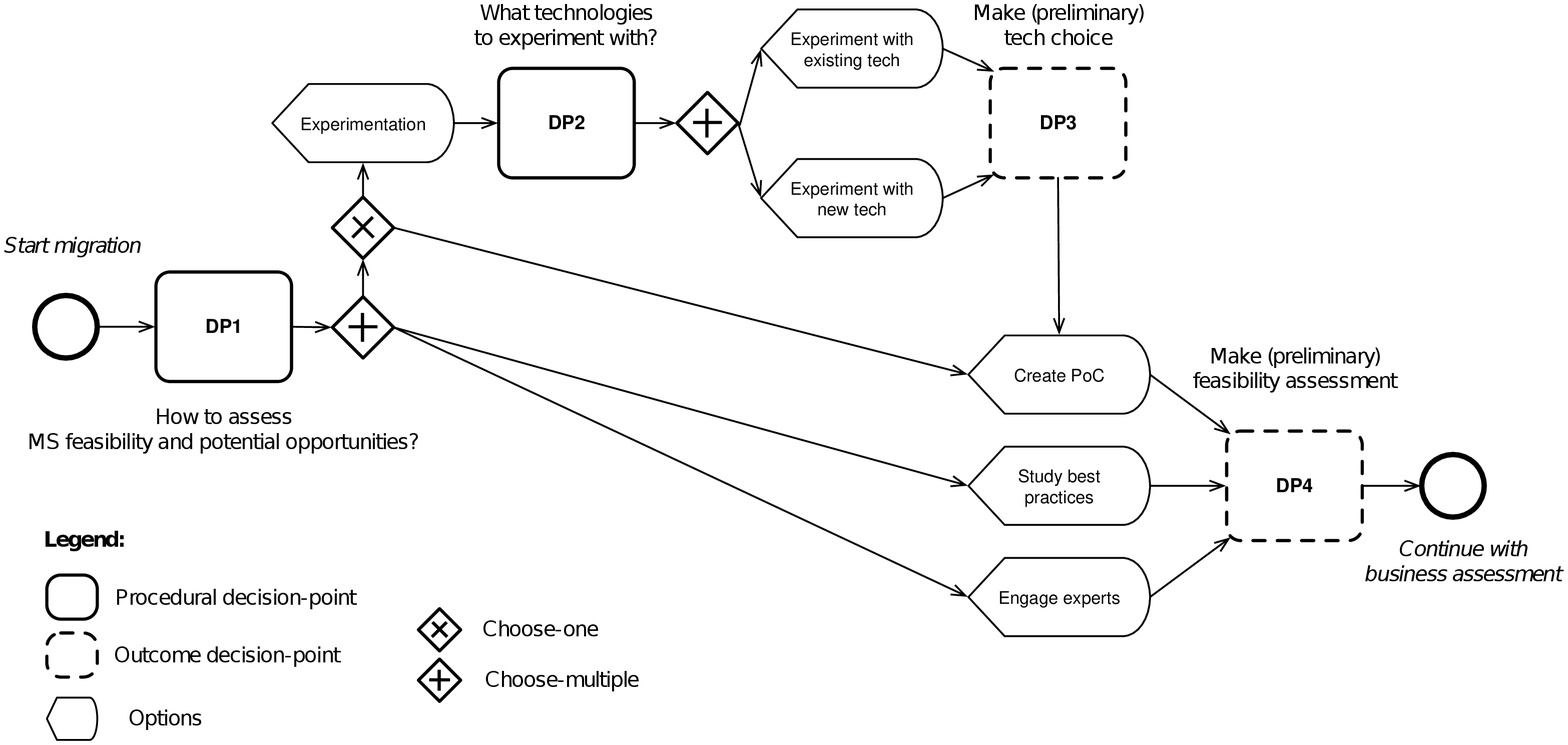}
  \caption{Technical Feasibility and Exploration of Opportunities}
  \label{fig:organizational}
\end{figure}

The initial part of the migration's decision-making process (depicted in Figure~\ref{fig:organizational}) is primarily about exploring the (technical or business) potential of microservices in the organization.

The first decision-point is \emph{how to assess microservice feasibility and potential opportunities or threats} (DP1).
Our interviewees describe four options for this assessment (experimentation, building proof-of-concepts or PoCs, studying best practices, and engaging with internal or external experts). In practice, many companies will opt for a combination of multiple or all of these options.
These options broadly fall in two groups. On one hand, highly practical and context-specific, with experimentation and building PoCs. On the other hand, theoretical and conceptual, with studying best practices and engaging with experts. In addition, experimentation and learning from best practices are based on knowledge that middle management and operational personnel obtain, whereas launching a PoC and hiring a specialist are based on practical experiences.

Experimentation is typically done early to gain understanding of the capabilities and limitations of concepts and technologies, and to assess how they apply to the specific organization, their projects, and products.
If a company chooses to rely on experimentation as an outcome of DP1, they face another decision, namely \emph{which technologies to experiment with} (DP2). Broadly, two options present themselves: (1) focusing on the technologies that the company already uses elsewhere and has experience in (experimenting with existing tech), (2) exploring new technologies and tools.
At the end of this activity, an outcome decision-point needs to be made, namely \emph{which of the technology stack(s) should be selected} -- at least for the time being (DP3). Worth noting is that experimentation might also be repeated later in the process, in the case of needing to evaluate more new tools or technologies.
The selected technologies are then commonly the basis of one or multiple larger PoCs. 
\vspace{-0.5em}
\interviewquote{we had a small PoC [...] and then step-by-step grow and scale up our project}{I7}

Some companies (e.g., the employers of I1 and I16) elect to skip this activity of experimentation, as the technologies to use are pre-decided (e.g., the company has a technology stack that it does not want to deviate from). In these cases, companies jump straight to building microservice PoCs.
Creating a PoC entails building a minimum viable version of the new system, that is then evaluated with customers or users. 
Additionally, it is sometimes used to demonstrate to management how the architecture will look like.  
\interviewquote{it's kind of an investment, because initially you have to spend some money in order to do the experiments, and then you can gain the knowledge}{I7}

In addition, organizations use more theoretical alternatives of building up microservice knowledge. One such option is the study of best practices, reports, blogs and courses that are available for self-learning. Alternatively, another way to learn is through following the documentation of frameworks dedicated for microservices (e.g. Quarkus, lagoon).
Essentially, middle management and operational personnel learn by themselves about the technology. Subsequently, they act on this knowledge and transfer it to the rest of the organization.

Finally, organizations often hire specialists (or re-allocate from other parts of the company) that can transmit their knowledge and accelerate the learning process of managers and team members.
For example, organizations hire experienced individuals that went through migrations many times in the past or have extensive experience on a specific framework/tool.
        \interviewquote{So we have assigned architects from the different providers that we interact with and take advice from on how we could use what on the newest services}{I9}

\subsubsection{Constructing the migration's business case.}
After technical feasibility is established, the next phase is to answer the question whether a migration also has business value, in order to engage top management and other stakeholders in middle-management that were not informed before (e.g. from marketing and sales).

\begin{figure}[h!]
  \centering
  \includegraphics[width=\linewidth]{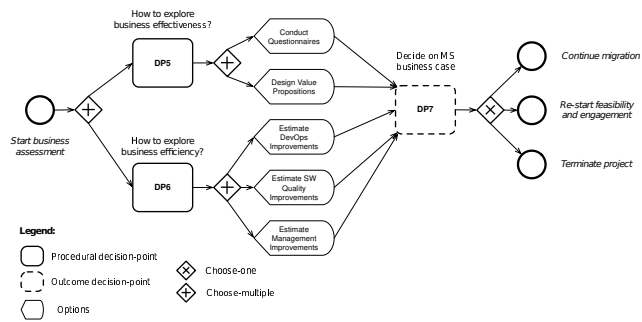}
  \caption{Constructing the migration's business cases}
  \label{fig:organizational2}
\end{figure}

As indicated in Figure~\ref{fig:organizational2}, the organizations represented in our study explore the business value of migrations along two separate axis, captured in DP5 (\emph{how to explore business effectiveness}) and DP6 (\emph{how to explore business efficiency}). Business effectiveness is about increasing the value that an organization delivers to its customers, typically through novel types of value-adding business services \cite{Venkatraman1986}. 
Business efficiency is about optimizing the ways in which an organization delivers value to customers and, thus, the operations in which value is created \cite{Venkatraman1999}.
These decision-points rely on the information and knowledge collected when checking on feasibility and exploring opportunities. Predominantly they serve as input for the construction and evaluation of the migration business case. It is worth noting that these decision-points emerged mainly from the more senior engineers interviewed (e.g. I5, I10, I13, I16, I19).

In DP5 (business effectiveness) organizations investigate how they will deliver more value to customers through their products or services.
For example, I16 describe how they are able to aggregate features more flexibly and deliver a larger variety of services that are customized for their clients' needs. 
I16 also mentions how it allowed them to sell more expensive or exclusive services, and ultimately increase profitability.
Additionally, I5 and I11 mention how it becomes possible to reach customers that before would not be reachable, by delivering smaller offerings that are easier to sell, with a subset of features, or just API calls. 
Finally, according to both I3 and I5, an important objective was making the sales process easier to clients. On one hand, smaller services act as "poaching" propositions that increase brand awareness and attract clients, sometimes through freemium models. On the other hand, smaller services reduce the barrier to closing a project with a customer.
\interviewquote{whenever you talk with a client about a huge system, they are going to think long time about it [...] but whenever you give them some microservices, you are sure that this is the only part they need}{I5} 

We find that common ways to identify these opportunities are typically to (1) conducting customer questionnaires and (2) designing / launching new value propositions. 
Conducting questionnaires can help adopt a customer-driven approach to the migration.
Our interviewees report that with questionnaires they establish communication with customers and involve them in development.
This can bring a better understanding of what delivers value, and, therefore, it helps in designing new value propositions.
Trying new modes of delivering value showcases to customers the value that they also gain from the migration. 
This enables the system's provider to sometimes even co-fund the migration with clients.
    \interviewquote{we start to find the client who is more interested [...] because for sure, for them to will become faster for maintaining the database and less time for testing as well.}{I5}

Another interesting point is that microservices can enable smaller offerings that are easier to sell since this allows providing cheaper products or services with a subset of features, thus reaching smaller organizations were not targeted before.
    \interviewquote{It was one huge system and I could not say I'm going to deliver it in one or two days. But when I do microservices, if you are a small organization, which you need to use my system, I can do it}{I5}

DP6 (business efficiency) is about the ways in which the new MSA will optimize the operations of the organization. 
Therefore, in this decision we observed options in which different bottlenecks were identified from the migration cases of this study and how they were improved. 
Three broad classes of possible improvements emerged in our study, some or all of which can be explored in this phase: improvements in operations and DevOps (e.g., more efficiency in systems operation), improvements in software quality (e.g., better performance at scale), and improvements in management processes (e.g., through easier recruitment of highly skilled individuals).

The information from all options that the organization decided to explore in DP5 and DP6 feed into the outcome decision-point DP7 (\emph{deciding on the microservices business case}). Here, one or multiple business cases are constructed based on all information collected so far. The business case(s) facilitate the deliberation between stakeholders, and are used for comprehending and aligning on both the business and technical dimension of the migration. 
Three alternatives present themselves as outcome of this business case construction and assessment: either to continue with the migration as planned (leading to commencing the technical migration), termination of the project (if obstacles are identified that seem not possible to overcome), or to revisit the decisions of technical feasibility and exploration of opportunities (e.g., if there appears to be potential, but additional information are needed, for instance by assessing a new technology or exploring additional value propositions).

\interviewquote{the best way is to say in the very beginning: Okay, this is how much it cost us now, [...] and then you say: Okay, if we move this out, then we would need that many resources [...] and costs should go down [...] and you can also scale up and down}{I10}
\vspace{-1em}

\begin{figure*}[h!]
  \centering
  \includegraphics[width=0.9\linewidth]{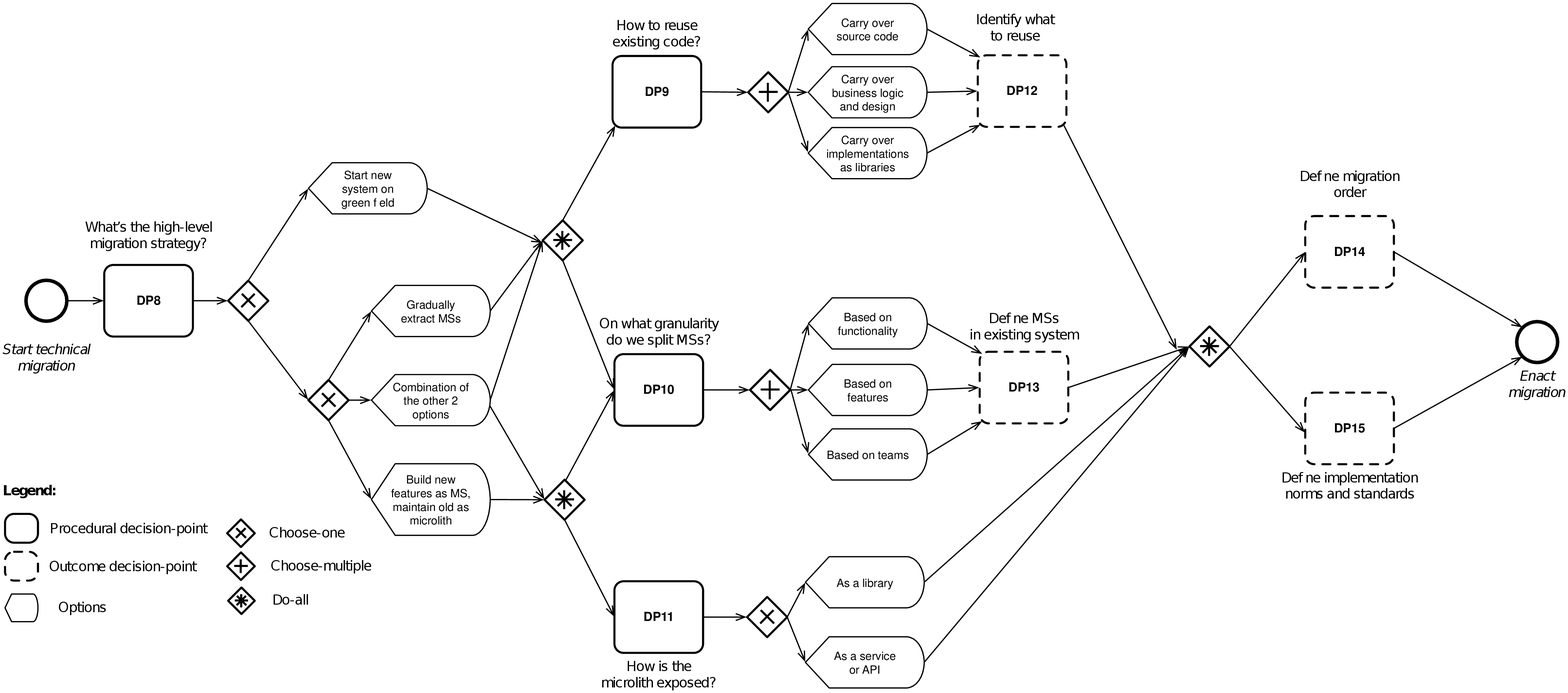}
  \caption{Technical decisions for migrating towards microservices}
  \label{fig:technical}
\end{figure*}

\subsection{Decisions on the Technical Dimension}
\label{sec:tech}

Once the business case for migrating towards microservices is made, the organization engages the technical migration. The decisions of the technical migration are summarized in Figure~\ref{fig:technical}. The first technical decision-point is DP8 (\emph{what is the high-level migration strategy}).
The migration cases we investigated revealed three alternative approaches to migrating a system towards microservices.

One alternative is to start the migration with architecting and developing a new system on the green field (from scratch).
This means that the monolith is not altered in any way and a new system is developed entirely from the beginning.
This case includes radical refactoring activities and usually leads to simultaneously developing and operating two systems until the microservice system is ready and the monolith can be turned off. 

On the other hand, the options are to gradually decomposing the monolith, retaining the monolith and building all new features as microservices or a combination of decomposing and building new features. 
These options result in a hybrid architecture, in which a small monolith (a "microlith") lives within the MSA.
The first option requires the extraction of microservices until the monolith is decomposed. This approach is about repeatedly extracting from existing functionality pieces of code that can be individual microservices, until the entire system is decomposed.
The second option is developing every new feature as a microservice and then operating a hybrid architecture until the old monolith phases out.

Finally, we find that a common choice among our interviewees is to combine gradual extraction with the retention of a "microlithic" kernel of old features. Those are for example features that are difficult or impossible to extract as microservies.
In this case, all new features are developed with microservices principles in mind and at the same time some features are extracted gradually, until the application reaches a level of granularity that is considered sufficient.
Our interviewees frequently mentioned that the perfect MSA cannot be achieved and the required mindset is to accept this and constantly improve the system.
\interviewquote{I don't think it's black and white where before we had a monolith and now we have a complete MSA. It's more like a gradual process, that might never be completed}{I5}

Once the high-level migration strategy is decided, organizations now reach decision-point DP9, and need to assess \emph{how to reuse existing code} (if anything shall be re-used at all, and if the old system is not simply re-used in it's entirety as a "microlith" using the third option from above).
Three different options for re-use have emerged in our study (which can evidently be combined). Re-using source code (e.g., by copying-and-pasting useful code), carrying over business logic and software designs that business analysts developed from requirements, or by encapsulating key functionality in libraries, which are then imported in the new microservices. It is worth noting that even organizations that started from scratch often reused some artifacts in that manner, though usually not source code.
This then leads to outcome decision-point DP12, which requires the organization to \emph{identify what concretely shall be reused} from the previous monolithic system.

Organizations that decide to develop all new functionality as microservices and leave the existing system as a "microlith" within the new architecture face a different decision-point (DP11), and need to decide \emph{how to expose the "microlith"}. The two observed choices to this end entail exposing the "microlith" as a library (similar to the re-use of other, smaller, code elements) or to host the "microlith" as a service or API. It is worth noting that a "microlith" is exposed in the investigated cases using a shell API.
\interviewquote{First step was to take the back-end as a whole, as one piece out of the front-end and connect them with one big library that is imported in the UI. And then we built on API around it and that's where we could have a back-end and the front end.}{I11}

Independently of their choice in DP8, an organization now also needs to decide \emph{on which granularity microservices shall be split} (DP10). For organizations starting on a green field or which gradually extract microservices, this decision applies to the services representing existing code. However, even if the old system shall be retained as a "microlith" this decision needs to be made for new services.
The first option that emerged from our interviews is to split services based on the functionality of the microservice, i.e., based on functions (for example sorting) or architectural layers (for example front-end). 
The second option is to split microservices based on features, i.e., designing microservices that are small end-to-end applications that go through the entire development stack. 
Finally, the third emerging option is to align the service structure with existing team structures. 
For example, if teams are in different geographic locations and time-zones and work independently on different tasks, the microservices are designed to accommodate this particularity.
Finally, a rule of thumb we observed is achieving the required level of granularity by keeping a microservice independent as long as the code required for communication is smaller than the code required to implement the microservice.
\interviewquote{There's always an amount of code and logic needed, just to get a service up and running [...] and if it is bigger than the actual business logic part of the service, then [...] it is too small and I figure out where it can fit instead.}{I15}

At this point in the migration the organization has made most planning-related decisions. What is still left to do are three outcome decision-points, where the organization takes in all decisions and lessons learned so far to decide on the actual structure of their new system. These decisions start with DP13 (\emph{define what microservices the new system should contain}). 
Then in DP14, (\emph{define the order of migration}) we observed that engineers have to decide which migration activities will take place in which chronological order. We observe from the interviews six main activities that constitute a migration - splitting the back-end, the database, the front-end, DevOps, microservices communications and finally reorganizing teams.

Finally, DP15 (\emph{define new implementation norms and standards}) 
allows to establish the governance for scaling the migration to cover most parts of the system.
\interviewquote{For it to make sense, you want it to scale across the organization and across the enterprise. And then you need to have kind of a strong governance within that organization}{I13}

Furthermore, these final decisions entail also concerns that interviewees thought are important to address early in the migration, since many migration cases showed evidence that neglecting or postponing some aspects entail risks.
\interviewquote{There were some things that should be said from the very beginning in order to avoid later additional efforts.}{I2}
For example, I7 mentions security and how it became a challenge to address later on if neglected in the migration.
\interviewquote{security is neglected from the start, making it hard to add it later on}{I17}
Another example is how not re-developing logging mechanisms and exception or error handling leads to costly future work.
\interviewquote{We started in the beginning, doing the development without deciding what the type of our logs will be, how we’ll do the exception handling [...] and now someone should go back to all the development that we did and put logs and put the correct errors and exceptions.}{I2}

\subsection{Decisions on the Organizational Dimension}
\begin{figure*}[h!]
  \centering
  \includegraphics[width=0.9\linewidth]{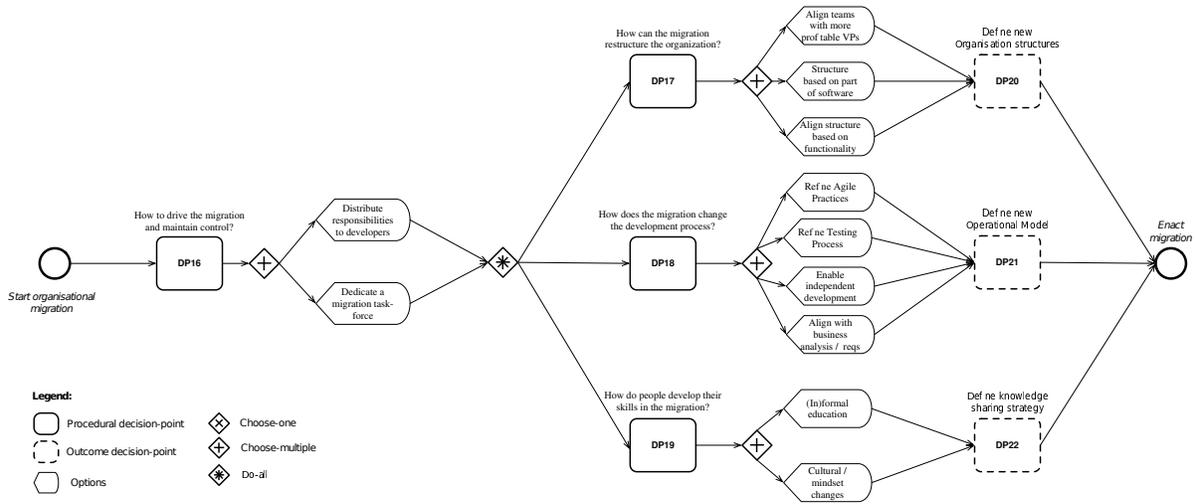}
  \caption{Restructuring the organizational structures and operations}
  \label{fig:org-restruct}
\end{figure*}

In addition to the business and technical designs discussed so far, a migration also entails a set of decisions that are taken to restructure the organization as well as the organization's operations.
The objective is typically to lead operations in the direction of achieving the estimated benefits from the business case.
Another objective is to align business with the new capabilities and requirements of the new technology, that come from the decisions described in Section \ref{sec:tech}.
We identify the constituent organizational decision-points, as summarized in Figure \ref{fig:org-restruct}. 

The first organizational decision-point is DP16 (\emph{how to drive the migration and maintain control}).
A migration typically takes a significant amount of time to complete, and the studied organizations use two options to maintain a stable track towards the new architecture. 
One option observed is to distribute responsibilities to developers.
The rationale is usually to distribute the migration activities in the entire organization. Hence, organizations avoid the risk of having isolated knowledge for the migration, that is not systematized or validated throughout the entire organization.
The other option is to dedicate a team specifically responsible for the migration. This can eliminate the perception that the migration is a project on the side, since a specialized team makes the project a main contributor to the organization's strategy. 
The rationale is also to accelerate the learning curve of the organization on MSA, by dedicating a specialized team to the new technologies.
    \interviewquote{Some team needs to work on the new product and you need to allocate some people and to invest with them}{I4}
    
Evidently, these two options also relate on how to maintain a mutual understanding and alignment between stakeholders about the change taking place. 
Specifically, this decision-point indicates the ways that consensus is built and communicated between stakeholders. 
This consensus can either be established top-down (through management decisions, e.g., in case of I12) or bottom-up (through extensive team discussion and deliberation, e.g., I2 or I3). Once it is clear how the migration will take place, organizations reach decision-point DP17 (\emph{how the organization is getting restructured}), for which three options emerged.

The first option is to align the teams' structures with the most profitable value proposition. For example, I5 mentions that if at a given point of time there is a need for engineers in a very profitable project, then people from different teams are assigned to it. I5 also mentions that this dynamism happens quite regularly and that decisions are made with an eye on immediate customer value.
\interviewquote{We started to present and see what is easier for them, especially from customer services, [...] if there's something that they [the customers] are more interested in.}{I5}

The second option is to align team-structures based on parts of the software developed. For example, I11 distinguishes front-end microservices from back-end microservices, having teams assigned to each part.
Finally, there is the third option with a way of organizing roles, responsibilities and ownership of services based on the functionality and the features developed.
    \interviewquote{break the team into smaller parts and say you and you will support this service, you and you will support the other services”}{I3}

Also, organizations reach the decision-point DP18 (\emph{how does the development process change}). 
Organizations optimize operations by simplifying their process, for example through enabling parallel development in parts of the code-base or with faster identification of issues.
We identify four options of altering the development process. 
One way is through refining the agile practices adopted from the organization.
Another option is through refining the testing process.
Additionally, another option way of process change is enabling the independent development from developers.
Finally, an identified option is aligning with business analysis and ways of extracting or communicating requirements throughout the process.
    \interviewquote{the guys that were writing the functional specs, you know the business requirements basically, they have to somehow map their ideas and their designs to specific microservices somehow}{I3}

The next decision-point is DP19, or \emph{how do people develop their skills and knowledge in migrations?}
Many interviewees mentioned cultural changes that need to take place and the learning curve required to take in order to succeed in the migration. 
\interviewquote{People needed to educate and train a lot. So, the company invested on that one as well, to train people}{I3}
We identify two options for DP19, coming from the fact that migrations change the expectations on engineers and management in terms of knowledge.
The first option is to provide (in)formal education of the new technologies, tools and frameworks.
Most importantly, the second option is giving time to experience the new architecture and develop knowledge in terms of learning a new paradigm of development and transforming the mindset and culture accordingly.
    \interviewquote{Also, maybe change of the mindset of some people as well. Both design people and developers}{I3}

Finally, the decision-making process of the organizational and operational restructuring concludes with three outcome decision-points. In DP20 organizations \emph{define the new organizational structures} and in DP21 \emph{define the new operational model}. DP22 is to \emph{define the knowledge sharing strategy of the organization} and how skills will be propagated across the organization.

\section{Discussion}

Decision-making typically happens in an individual level, team level and organizational level. The focus of our findings is on the complex decisions that take place on a team level and an organizational level. That means that we see migrations at a scale of the entire software development organization and not only isolated solutions. 

\emph{Multidimensionality of migrations:}
First, our findings show that migrations towards microservices are not only about technological change, but also other dimensions.
For example, microservices often facilitate potential cross-team collaboration and autonomy of teams. This, in combination with how teams' structures change, shows us how microservices are not only a software architecture but also an enabler of organizational restructuring.
The dimensions we identify are in line with socio-technical systems engineering, containing organizational structures (Section 4.3), processes (contents of decision-making processes), technologies (Section 4.2) and people (stakeholders) \cite{Baxter2011}. Additionally, we demonstrate how to achieve comprehensive strategic alignment to leverage microservices for delivering strategic value  \cite{Venkatraman1999}. Specifically, this takes place by including the aspect of business effectiveness \cite{Venkatraman1986} in decision-making and considering how it will increase the delivery of value to customers. For example, through new sales models or from aggregating different combinations of services. The aspect of business effectiveness complements existing work on the economics of microservices that demonstrate how the architecture improves efficiency \cite{singleton2016}.

\emph{Human-centered approach in microservices migrations:}
Adopting a value-based approach can help to accommodate the needs of all stakeholders and achieve stronger engagement.
Decision-making mechanisms of migrations are often not centralized, spanning from board-room level, to operational level.
Some of the most complex and under-studied aspects during migrations are the dynamics between stakeholders in decision-making.
We address this gap by aggregating a set of decisions that organizations make during migrations. Our theory contributes by providing a process for understanding the different perspectives and incentives that usually prevail among different stakeholders. 

We evolve existing knowledge of decision-making in different stakeholders views \cite{Kruchten2009}, by demonstrating in detail these decisions and the point in which they are made.
For example, to ensure the engagement of key stakeholders from within the organization it is essential to consider both their business-oriented or technical-oriented backgrounds and roles. 
The demonstrated value ranges from aspects like increasing profitability, to potentially tackling talent scarcity (due to polyglot nature) and fast on-boarding of new employees (small and easily comprehensible microservices). 

\emph{A pragmatic view on microservices migrations:} 
In addition, a common view of all interviews was that microservices are not a silver bullet, and that their benefits need to be investigated critically, as early as possible.
Otherwise, organizations might unjustly choose to migrate in cases that do not fit with the architecture.
This can lead to failed migrations and a negative attitude towards the technology, despite the fact that in different cases from those attempted the architecture might be a better fit. 
In addition, the technical overhead that can come with microservices is not always worthy of the needed effort. Even though a MSA has many benefits, organizations should still consider alternatives like keeping a monolith, purchasing a solution, or outsourcing development. 

Moreover, our decision-making processes are modeled as linear, mainly for visualization purposes and showing the points in which decisions are made. Decision-making processes can repeat in different iterations when migrating. 
However, a migration project eventually has a structure and a sequence of events, that we show through Decision-Points. 
The core of microservices migration projects is to maintain a capacity of continuously unravelling architectural design inefficiencies. Then, actively eliminating them, with decision-making that moves the software development organization forward. 

Furthermore, early decisions can sometimes influence also subsequent decisions.
For example, the different ways in which an organization evaluates and explores the value from microservices can influence the course that the migration will eventually take.
Decisions of exploring the potential of microservices with experimentation or a PoC can help to gain more accurate estimates of tasks, capabilities and limitations. Hence, planning can happen with a better understanding of the time needed and the potential risks.

\emph{Practical \& Theoretical Relevance:}
We envision practitioners to use our findings complementary to the available best practices, through the decisions they are required to make. 
Our decision-making processes support practitioners in making more educated decisions that are better informed, considering different potential alternatives and adding to existing decision support frameworks \cite{AUER2021106600}. Best practices can provide guidance on how to setup a microservices architecture, but cannot describe how practitioners decide on taking the suggested actions, or what alternatives they consider. 
Some best practices can be taken with a grain of salt, since there are so many different potential paths to migrating for different situations. 
Also, our evidences show that there is a large spectrum of moderately adopting some practices. The different options can all be made to a lesser or greater extent, depending on many situational factors. 

This work paves the way on empirically backing and understanding microservices migrations and their underlying decisions. 
There is significant research on the outcomes or contents of decisions \cite{Hassan2020}, and we build on it. 
Some of our observed decisions that resulted are in line with existing literature \cite{SOLDANI2018215}. For example, choosing a migration strategy between starting from greenfield or gradually migrating are similar to existing industry practices \cite{newman2019monolith}. 
However, more investigation is needed on the decision-making processes and the approaches to resonate about alternative choices. 
Hence, the focus of future work needs to not only provide knowledge about the outcome of decisions, but rather on identifying the reasoning behind those decisions. %This rationale can be propagated to other outcomes of decisions in migrations.
Future research can also target the evaluation of such detailed decision-making processes.

\section{Conclusion}

In this paper, we investigate how software engineers and organizations go through migrations towards microservices. Specifically, we obtain an understanding of the different decisions involved in migrations along with their interplay.
The 16 different cases of migrations towards microservices that we investigate in this study reveal to us details on the different dimensions of decision-making during migrations.
On one hand, a set of decisions take place in order to prove the technical feasibility of a microservices-based architecture.
On the other hand, there are decisions regarding the driving forces for pursuing a migration on the first place and how engagement is achieved. This shows how to create the required "buy-in" in order to make a microservices migration get in grips across the organization.
In addition, we see decision-points that take place on the technical dimension and how these resonate with decision-points that take place in an organizational and operational level. Importantly, we see how these dimensions are complementing each other to motivate migrations more strongly and proceed at changing the software architecture with better awareness. 

\begin{acks}
This work was done as part of TrAF-Cloud funded by the Swedish innovation agency VINNOVA (proj.no. 2018-05010). We would also like to express our appreciation to all the participants of the study for their time, support and valuable insights during the interviews. 
\end{acks}

\bibliographystyle{ACM-Reference-Format}
\bibliography{references}

\end{document}